\documentclass[12pt]{article}

\usepackage{graphicx,rotating,epsfig}  
\usepackage{multirow}
\newlength{\digitwidth} 

\newcommand{\mlab}[1]%
{\mbox{}\marginpar{\raggedright\hspace{0pt}\footnotesize #1}}

\begin{document}

\begingroup
\thispagestyle{empty}
\baselineskip=14pt
\parskip 0pt plus 5pt

\begin{center}
\large EUROPEAN LABORATORY FOR PARTICLE PHYSICS
\end{center}

\bigskip
\begin{flushright}
CERN-PH-EP\,/\,2006--019\\
7 June 2006
\end{flushright}

\bigskip\bigskip
\begin{center}
\Large\bf
$\psi'$ production in Pb-Pb collisions\\ at 158 GeV/nucleon
\end{center}

\bigskip\bigskip
\begin{center}
\small
\end{center}

\begin{center}
\emph{NA50 Collaboration}
\bigskip
\end{center}

\begin{center}
B.~Alessandro$^{10}$,
C.~Alexa$^{3}$,
R.~Arnaldi$^{10}$,
M.~Atayan$^{12}$,
S.~Beol\`e$^{10}$,
V.~Boldea$^{3}$,
P.~Bordalo$^{6,a}$,
G.~Borges$^{6}$,
J.~Castor$^{2}$,
B.~Chaurand$^{9}$,
B.~Cheynis$^{11}$,
E.~Chiavassa$^{10}$,
C.~Cical\`o$^{4}$,
M.P.~Comets$^{8}$,
S.~Constantinescu$^{3}$,
P.~Cortese$^{1}$,
A.~De~Falco$^{4}$,
N.~De~Marco$^{10}$,
G.~Dellacasa$^{1}$,
A.~Devaux$^{2}$,
S.~Dita$^{3}$,
O.~Drapier$^{9}$,
J.~Fargeix$^{2}$,
P.~Force$^{2}$,
M.~Gallio$^{10}$,
C.~Gerschel$^{8}$,
P.~Giubellino$^{10}$,
M.B.~Golubeva$^{7}$,
M.~Gonin$^{9}$,
A.~Grigoryan$^{12}$,
J.Y.~Grossiord$^{11}$,
F.F.~Guber$^{7}$,
A.~Guichard$^{11}$,
H.~Gulkanyan$^{12}$,
M.~Idzik$^{10,b}$,
D.~Jouan$^{8}$,
T.L.~Karavicheva$^{7}$,
L.~Kluberg$^{9}$,
A.B.~Kurepin$^{7}$,
Y.~Le~Bornec$^{8}$,
C.~Louren\c co$^{5}$,
M.~Mac~Cormick$^{8}$,
A.~Marzari-Chiesa$^{10}$,
M.~Masera$^{10}$,
A.~Masoni$^{4}$,
M.~Monteno$^{10}$,
A.~Musso$^{10}$,
P.~Petiau$^{9}$,
A.~Piccotti$^{10}$,
J.R.~Pizzi$^{11,d}$,
F.~Prino$^{10}$,
G.~Puddu$^{4}$,
C.~Quintans$^{6}$,
L.~Ramello$^{1}$,
S.~Ramos$^{6,a}$,
L.~Riccati$^{10}$,
A.~Romana$^{9,d}$,
H.~Santos$^{6}$,
P.~Saturnini$^{2}$,
E.~Scomparin$^{10}$,
S.~Serci$^{4}$,
R.~Shahoyan$^{6,c}$,
M.~Sitta$^{1}$,
P.~Sonderegger$^{5,a}$,
X.~Tarrago$^{8}$,
N.S.~Topilskaya$^{7}$,
G.L.~Usai$^{4}$,
E.~Vercellin$^{10}$,
L.~Villatte$^{8}$,
N.~Willis$^{8}$\\
\end{center}

\begin{abstract}
$\psi'$ production is studied in Pb-Pb collisions at 158 GeV/c per nucleon incident momentum. Absolute cross-sections are measured
and production rates are investigated  as a function of the centrality of the collision. The results are compared with those obtained for lighter colliding systems and also for the J/$\psi$ meson produced under identical conditions. \\

\begin{center}
{\it\normalsize Accepted for publication in European Physics Journal C}\\
\end{center}
\end{abstract}

\newpage
\thispagestyle{empty}

\noindent
$^{~1}$ Universit\'a del Piemonte Orientale, Alessandria and INFN-Torino, Italy\\
$^{~2}$ LPC, Univ. Blaise Pascal and CNRS-IN2P3, Aubi\`ere, France\\
$^{~3}$ IFA, Bucharest, Romania\\
$^{~4}$ Universit\`a di Cagliari/INFN, Cagliari, Italy\\
$^{~5}$ CERN, Geneva, Switzerland\\
$^{~6}$ LIP, Lisbon, Portugal\\
$^{~7}$ INR, Moscow, Russia\\
$^{~8}$ IPN, Univ. de Paris-Sud and CNRS-IN2P3, Orsay, France\\
$^{~9}$ LLR, Ecole Polytechnique and CNRS-IN2P3, Palaiseau, France\\
$^{10}$ Universit\`a di Torino/INFN, Torino, Italy\\
$^{11}$ IPN, Univ. Claude Bernard Lyon-I and CNRS-IN2P3, Villeurbanne, France\\
$^{12}$ YerPhI, Yerevan, Armenia\\

\noindent
a) also at IST, Universidade T\'ecnica de Lisboa, Lisbon, Portugal\\
b) also at Faculty of Physics and Nuclear Techniques, AGH University of Science and Technology, Cracow, Poland\\
c) on leave of absence of YerPhI, Yerevan, Armenia\\
d) Deceased\\

\endgroup

{\it Dedicated to the memory of our colleagues Albert Romana and Jean Ren\'{e} Pizzi, whose  
untimely death occurred while finalizing this article.}




\newpage
\pagenumbering{arabic}
\setcounter{page}{1} 

\section{Introduction}
The production of charmonium states in ultrarelativistic heavy ion collisions 
has been investigated since the mid 80s by the NA38 Collaboration~\cite{na38}, followed in the 90s by NA50~\cite{na50_PbPb2}. The motivation is to search for the phase transition of nuclear matter from its normal state to a deconfined quark-gluon plasma,
as predicted to occur, under extreme energy densities or temperatures, by non-perturbative QCD. In such a deconfined matter, charmonia states, in particular the J/$\psi$ vector-meson, have been predicted to be suppressed by Debye colour screening~\cite{MatSatz}. Indeed, the NA50 experiment has shown that J/$\psi$ production in Pb-Pb collisions is significantly suppressed with respect to the expectations derived from the rates measured in proton, oxygen and sulphur-induced reactions ~\cite{na50_pA1,na50_pA2,na50_pA3}. The $\psi'$ is a more loosely bound state than the J/$\psi$, which opens the
possibility that it may dissociate at lower energy densities in connection
with critical scenarios, as predicted by recent lattice
calculations~\cite{Karsch_Ericeira, Alberico}, but also be sensitive to other mechanisms, as absorption in nuclear matter or dissociation by co-moving hadrons.
After the first results on $\psi$' production obtained by experiments NA38~\cite{na38_psip} and NA50~\cite{QM2004_psip}, the framework for a general understanding of charmonium dissociation due to critical transitions has recently opened up with new experimental results on 
J/$\psi$ suppression~\cite{psi_RHIC, psi_NA60}. 
The measurement of $\psi'$ production reported hereafter, performed
with identical analysis procedures as for the J/$\psi$,   
should  help and internally constrain the various  theoretical approaches which 
try to give a coherent overall picture of charmonium production and suppression.\\   
\indent
\section{Experimental setup, data selection and analysis method}
\label{Experimental setup, data selection and analysis method}
\begin{figure}[ht]
\centering
\includegraphics*[width=.65 \textwidth]{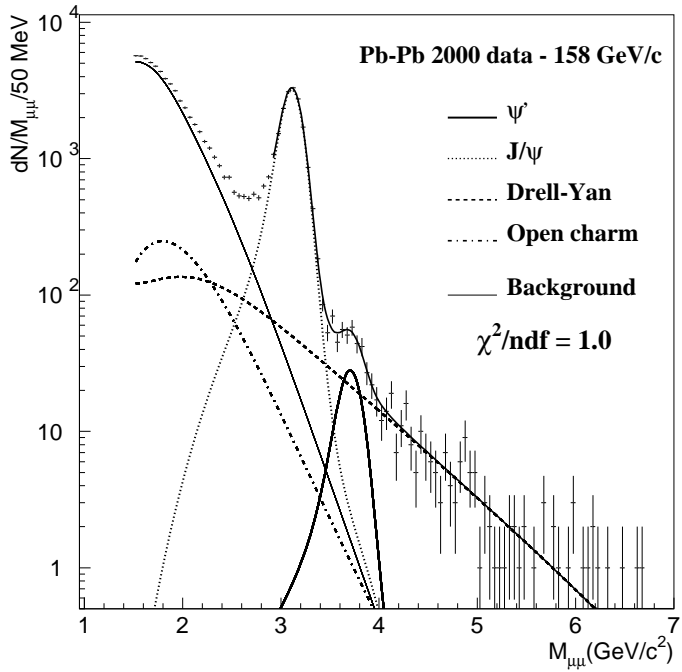}
\caption{The fitted opposite sign dimuon invariant mass spectrum, for
  mid--centrality Pb-Pb collisions.}
\label{fig:mass}
\end{figure}

NA50 is a fixed target experiment at the CERN-SPS optimized for
the detection of muon pairs produced in nucleus-nucleus collisions.
The main component of the apparatus is a muon spectrometer
made of
an air-core toroidal magnet surrounded by two sets of multi-wire proportional chambers and trigger hodoscopes.
An appropriate hadron absorber separates the spectrometer from the target itself, allowing the experiment to run with a high intensity incident beam and a moderate spectrometer illumination.     
Three independent detectors, located 
immediately
downstream from the target region, measure the centrality of the reactions --- 
a silicon strip multiplicity detector sampling the produced secondary charged particles, a Pb-scintillating fibers electromagnetic calorimeter measuring the neutral transverse energy $E_{T}$ and a very forward hadronic calorimeter (zero degree calorimeter), embedded in the hadron absorber,
which essentially measures the energy of the beam spectator nucleons in the
collision, $E_{ZDC}$. A beam hodoscope (BH), located 25\,m upstream from the target, allows to monitor and identify the incident beam and also provides an accurate time reference for trigger purposes.   
A detailed description of the NA50 apparatus can be found in Ref.~\cite{NA50_B410_327}.\\
\indent
This article is based only on the highest quality data samples collected by the experiment in years 1998 and 2000 when only one single target was used which, moreover, was placed in vacuum in year 2000. These features have efficiently solved problems affecting previous data collections such as secondary fragment reinteractions within the target assembly leading to a 
centrality smearing on one hand and Pb-air interactions simulating peripheral Pb-Pb collisions on the other. Furthermore, the selected data have benefited from an upgraded track reconstruction treatment based on improved software algorithms~\cite{na50_pA2, Ruben_tese}.
The data were collected with a typical beam intensity of 
1--1.4$\times$10$^7$ ions/s, over 4.8\,s bursts every
20\,s, at 158 GeV per nucleon. The single Pb target was 4\,mm thick in
2000 (3\,mm in 1998), equivalent to
10\% (7\%) of an interaction length.
Muon pairs 
are selected in the rapidity window $2.92\leq y_{lab}<3.92$ $(0 \leq y_{CM} < 1)$ and with a Collins-Soper angle $\mid {\rm cos}\theta_{CS}\mid<0.5$, resulting in an acceptance of around 14\%. On-target interactions are selected requiring the proper correlation between hits in the two planes of the multiplicity detector. 
This target selection procedure is 88\% efficient for $E_{T}> 5$~GeV and 100\% efficient
for $E_{T}> 22$~GeV.
It allows to unambiguously identify the most peripheral on-target interactions and,
therefore, has now been applied to the 
Pb-Pb data samples collected both in 1998 and in 2000.
Moreover, 
muons produced off-target are further excluded by a cut on the transverse
distance, in the middle target plane, between the muon track extrapolated
from the spectrometer and the beam axis.
Parasitic interactions occurring upstream from the target, mostly in the beam hodoscope, are rejected by a BH interaction detector and by anti-halo counters. 
Multiple piled-up interactions are discarded by a shape analysis of the signal from the electromagnetic calorimeter. 
Residual piled-up events are rejected by a 2$\sigma$ 
cut in the $E_{T}-E_{ZDC}$ correlation~\cite{na50_PbPb2}.\\
\indent
The opposite-sign dimuon invariant mass spectrum in the mass region of interest, 
as shown in Fig.~\ref{fig:mass}, results from five  different
contributions --- the resonances J/$\psi$ and $\psi'$, as well as the continuum formed by Drell-Yan, by
open charm semi-leptonic decays and by the {\sl unphysical} combinatorial background 
originating mostly from uncorrelated $\pi$ and $K$ decays. In order to disentangle the different contributions and to provide the
 necessary ingredients for the determination of cross-sections, a Monte-Carlo simulation technique is used. It allows to determine  both the acceptance of each physical process and the smearing of the corresponding detected muon pairs as they result from the actual experimental  detector and selection criteria imposed on the data. Each physical process is separately generated and its corresponding muon pairs are propagated through the detector exactly the way real muons do, which accounts, in particular, for multiple scattering and energy loss in the absorbers and for the limited acceptance of the apparatus. The resulting smearing is
particularly visible in the J/$\psi$ and $\psi'$ reconstructed shapes,
entirely due to the experimental resolution of the apparatus. 
The resonances are generated in rapidity with a Gaussian distribution 
of 0.6 units width. The transverse momentum used in the
simulation follows the parametrization of a thermal distribution, 
d$\sigma/dp_{T}=p_{T}M_{T}K_1(M_{T}/T)$, where $K_1$ is the modified Bessel function of the second kind and first order, in the transverse mass, with $T$=236 MeV. These parameters were tuned in order to reproduce the experimental spectra of the S-U collision system at 200 GeV/c~\cite{Carlos_tese} and they account very well for Pb-Pb interactions at 158 GeV/c~\cite{hs_thesis}. The Drell-Yan is evaluated with PYTHIA~\cite{PYTHIA}, using the GRV LO 94~\cite{grvlo94} set of parton density functions, 
with a minimum 4-momentum transfer and $x$ domain as required by the NA50 phase space window. At the Born level, dimuons are produced with zero $p_{T}$; to overcome this inadequate physical picture, a primordial $k_{T}$ inside the hadrons is considered, according to a
gaussian with 0.8 GeV/c width. The $D\overline{D}~$ mesons are also generated with PYTHIA, at leading order of QCD, using $m_{c}$ = 1.35 GeV/c$^2$ and a gaussian distributed $k_{T}$ with 1.0 GeV/c width~\cite{Capelli_tese}. The combinatorial background, free from any simulation, is completely determined both in shape and in amplitude from the measured like-sign pair
 distributions according to $N_{BG}^{+-}=2 \sqrt{N^{++}N^{--}}$. This method
 holds as long as the spectrometer acceptance does not depend on the muon charge, which is
 ensured by an appropriate selection of events.\\
\indent
 After the Monte-Carlo determination of the muon pair mass shapes of each physical
process, their corresponding amplitudes are obtained from a fit
 to the opposite sign dimuon invariant mass spectrum of the selected events. The background contribution,
 already established through the procedure explained above, is fixed.
  The $D\overline{D}~$ contribution, although tiny in the mass range of relevance for the $\psi'$ amplitude determination and, moreover, practically uncorrelated with it, is roughly
 estimated from a separate fit in the dimuon mass range 1.7-2.2 GeV/c$^2$. For this purpose, the  corresponding Drell-Yan is deduced from the extrapolation
 of the muon pairs sampled above 4.2 GeV/c$^2$, a mass region where Drell-Yan is the only
 contribution and is, therefore, unambiguously determined.
  The $\psi'$, J/$\psi$ and the whole Drell Yan amplitudes are obtained from a fit
 in the mass range 2.9-7.0 GeV/c$^2$, with both the combinatorial background and $D\overline{D}~$ contributions fixed. The set of parameters which maximize the log-likelihood function is found by the MINUIT package~\cite{minuit}, which also evaluates the function in the neighbourhood of the minimum, thus providing the parameter statistical uncertainties. One should notice the good $\chi^2$ of the fit shown in Fig.~\ref{fig:mass}. Details on the analysis method are given in~\cite{hs_thesis}.\\
\indent
\section{Absolute cross-sections}
\label{Absolute cross-sections}
The study of $\psi'$ production in Pb-Pb collisions implies special care with
the data treatment, in particular in what concerns the identification of systematic sources, since this resonance yields  
a weak signal due both to its small dimuon branching ratio
($B_{\mu^{+}\mu^{-}}$ = (7.3~$\pm$~0.8)~$\times$~10$^{-3}$~\cite{pdg})
and to its large suppression with increasing centrality of the collision. Another challenge is the extraction of the $\psi'$ signal itself from the several overlapping dimuon sources.

\begin{table}[t!] 
\centering
\begin{tabular}{|c|}
\hline\\[-4mm]
{\bf Detector efficiencies (\%)}\\[1mm]
\hline\\[-4mm] 
Dimuon trigger~~~~~~~~~~~~~~~~~~~~~~~~~~~~~~~~~~87 $\pm$ 1\\
Muon track reconstruction~~~~~~~~~~~~~~~~~~~94 $\pm$ 1\\
Beam hodoscope~~~~~~~~~~~~~~~~~~~~~~~~~~~~~~~~83 $\pm$ 1\\
Lifetime of DAQ~~~~~~~~~~~~~~~~~~~~~~~~~~~~~~~~96 $\pm$ 1\\
Target identification (5$<E_{T}<$22 GeV) ~87 $\pm$ 2\\[1mm]
~~~~~~~~~~~~~~~~~~~~~~~~~~~~~~ ($E_{T}>$22 GeV) ~$\approx$ 100\\[1mm]
\hline\\[-4mm]  

{\bf Selection cut losses (\%)}\\[1mm]
\hline\\[-4mm]
Interaction pile-up~~~~~~~~~~~~~~~~~~~~~~~15.8 $\pm$ 1\\
$E_{T}-E_{ZDC}$ correlation~~~~~~~~~~~~~~~~~16.3 $\pm$ 1\\
BH interaction~~~~~~~~~~~~~~~~~~~~~~~~~~~2.2 $\pm$ 0.5\\[1mm]
\hline\\[-4mm]

{\bf   Acceptances (\%)}\\[1mm]
\hline\\[-4mm]
~~~~~~$\psi'$~~~~~~~~~~~~~~~~~~~~~~~~~~~~~~~~~~~~14.8 $\pm$ 0.3 \\
~~~~J/$\psi$~~~~~~~~~~~~~~~~~~~~~~~~~~~~~~~~~~~12.5 $\pm$ 0.2 \\
Drell-Yan(2.9 $<M_{\mu\mu}<$ 4.5)~~~~~~~~13.8 $\pm$ 0.2  \\
Drell-Yan(4.2 $<M_{\mu\mu}<$ 7.0)~~~~~~~~17.8 $\pm$ 0.7  \\[1mm]
\hline

\end{tabular}
\caption{\label{tab:cor} Detector efficiencies, selection cut losses and integrated acceptances.}
\end{table}
\begin{table}[t!] 
\centering 
\begin{tabular}{|c|}
\hline\\[-4mm]
{\bf Cross-sections ($\mu$b) in $0 \leq y_{CM} < 1$ and $\mid \cos\theta_{CS}\mid<0.5$}\\[1mm]
\hline\\[-3mm]
Pb + Pb $\longrightarrow$ $\psi'~+~X$ = 0.136~$\pm$~0.013~$^{+0.010}_{-0.006}$\\ 
\\
Pb + Pb $\longrightarrow$ J/$\psi$$~+~X$ = 23.32~$\pm$~0.16~$\pm$~0.84\\
\\
Pb + Pb $\longrightarrow$ Drell-Yan(2.9 $<M_{\mu\mu}<$ 4.5) $~+~X$ = 1.37~$\pm$~0.07~$\pm$~0.05\\
\\
Pb + Pb $\longrightarrow$ Drell-Yan(4.2 $<M_{\mu\mu}<$ 7.0) $~+~X$ = 0.172~$\pm$~0.009~$\pm$~0.009\\[1mm]
\hline
\end{tabular}
\caption{\label{tab:cx} Inclusive cross-sections for $\psi'$ and J/$\psi$, multiplied by their branching ratios into $\mu^{+}\mu^{-}$, and for Drell-Yan, in the 4.2--7.0 GeV/c$^2$ mass range.} 
\end{table}

In order to measure absolute cross-sections, data collected in
the year 2000 were selected under strict criteria with respect to the
stability of the experimental conditions. 
After the data selection described in the previous section and after
correcting for the target identification inefficiency, about 900 $\psi'$ 
are left for further analysis. This absolute number of events must still be
corrected for various detector inefficiencies, selection cut losses and
acceptances in the NA50 phase space window, as detailed in Table~\ref{tab:cor}. The main systematical uncertainty contributing to the errors of the
acceptance values is the generation model assumed. In particular, if a Fermi
motion model of the nucleons inside the nucleus is taken into account, the
spectrometer acceptance of the resonances changes by $\sim$1.5\%. In the case of
Drell-Yan, different parton density functions lead to a 4\% difference in the acceptances.\\
\indent
Absolute cross-sections for $\psi'$ and J/$\psi$ in the dimuon channel, as well as for Drell-Yan in the mass windows of 2.9 $\le$ M$_{\mu\mu}$ $<$ 4.5 and 4.2 $\le$ M$_{\mu\mu}$ $<$ 7.0 GeV/c$^2$, are displayed in Table~\ref{tab:cx} within the kinematical ranges defined in section 2.
The $\psi'$ absolute cross-section is affected by a systematic
uncertainty of 7\%,  
obtained as the quadratic sum of all uncertainties considered,
namely the errors on the absolute normalizations, fit method, spectrometer acceptances and uncertainties associated with the Monte-Carlo inputs, like parton density functions and $c$-quark mass. The latter, although 
directly related to Drell-Yan and open charm theoretical uncertainties,
indirectly influence the $\psi'$ fitted normalization. The main contributions
to the uncertainty in the $\psi'$ yield are the
luminosity calculation, 4\%, and the chosen PDFs to generate Drell-Yan,
5.8\%, estimated
as the numerical change in the result when using
CTEQ4L~\cite{cteq} instead of GRVLO94. On the other hand, 
it has been verified that the more recent GRVLO98~\cite{grvlo98} 
induces a variation of less than 1.5\% on the cross-section measurement. The systematic uncertainty due to the fit method is 2.8\%.
The J/$\psi$ cross-section
obtained here, 23.32~$\pm$~0.16~$\pm$~0.84 $\mu$b, is 6.5\% higher than the one derived from the Pb-Pb data collected in year 1995,  21.9~$\pm$~0.2~$\pm$~1.6 $\mu$b, reported in Ref.~\cite{NA50_B410_327}. The difference results from the cumulative effects of the new J/$\psi$ experimental line shape~\cite{Ruben_tese} and the higher number of tracks reconstructed. Although the better description of the resonance tails would lead to a 3.5\% smaller cross-section, the new offline software algorithms~\cite{na50_pA2} deal much better with the high MWPC occupancy levels induced by the Pb-Pb interactions and result in a net \emph{increase} of the final cross-section value.\\
\begin{figure}[h]
\centering
\includegraphics*[width=.5 \textwidth]{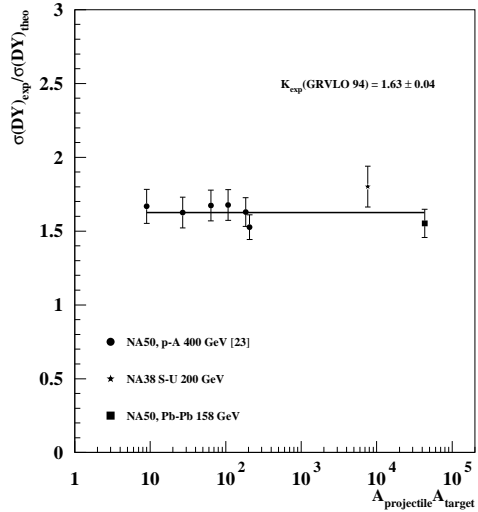}
\caption{The Drell-Yan $K_{exp}$ factor measured at 158--400 GeV/c beam
momenta in p-A and A-B collisions. GRV LO94 PDFs are used in
the theoretical calculations and in the extraction of the measured values.}
\label{fig:kdy}
\end{figure}
\indent
Figure~\ref{fig:kdy} shows the $K_{\rm exp}$ factor, the ratio between
the measured Drell-Yan cross-section and
the lowest order theoretical Drell-Yan cross-section. The
constant behaviour of this ratio from p-Be to Pb-Pb collisions shows
that the Drell-Yan process is proportional to the number of nucleon-nucleon collisions in the NA38/NA50 phase space window.
\indent
\section{Cross-section ratios versus centrality}
\label{Cross-section ratios versus centrality}
\begin{figure}[!ht]
\begin{center}
 \begin{minipage}[h]{.47\linewidth}
 \includegraphics[width=\linewidth]{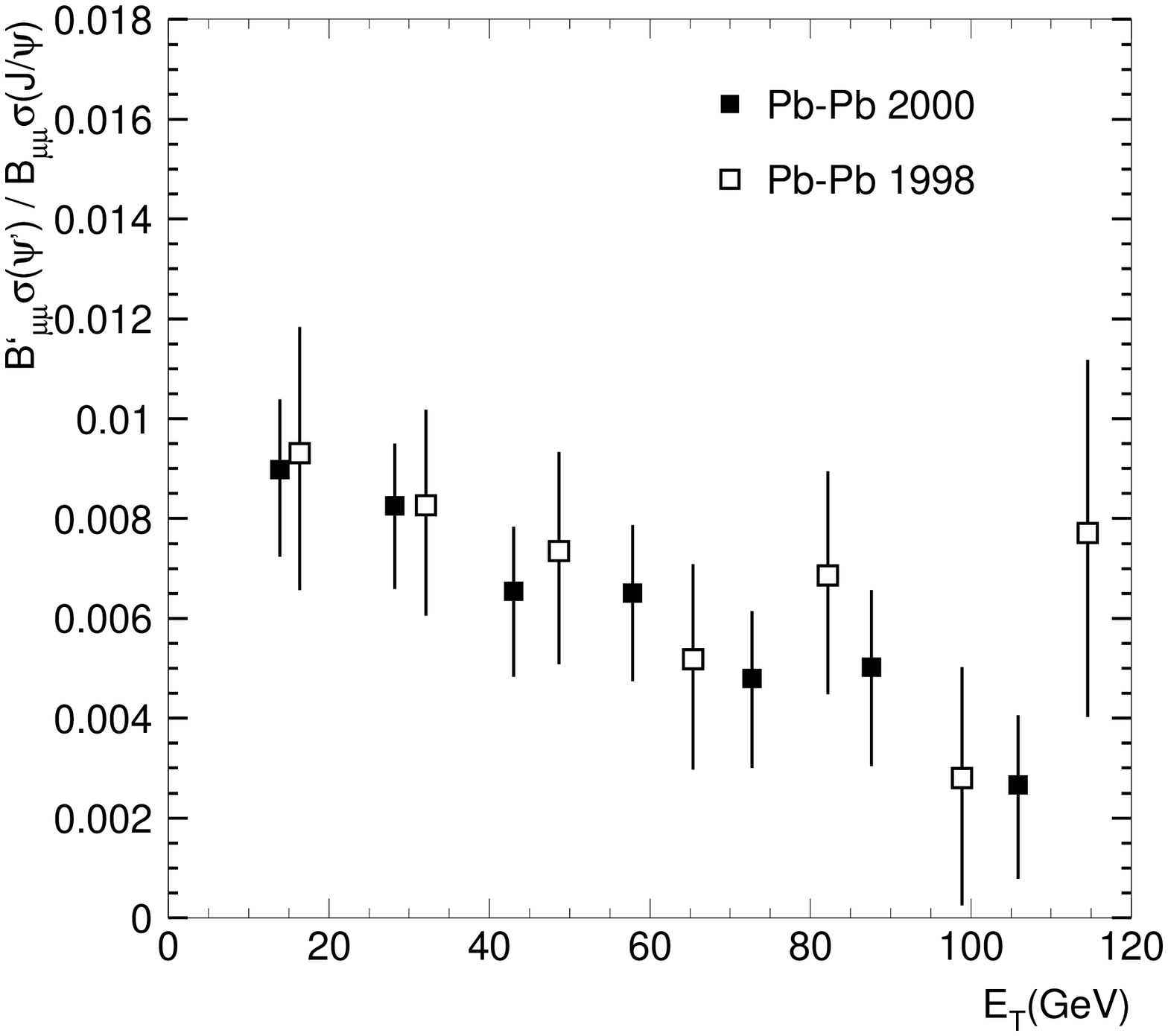}
 \end{minipage} 
\hspace*{6mm}
 \begin{minipage}[h]{.47\linewidth}
  \includegraphics[width=\linewidth]{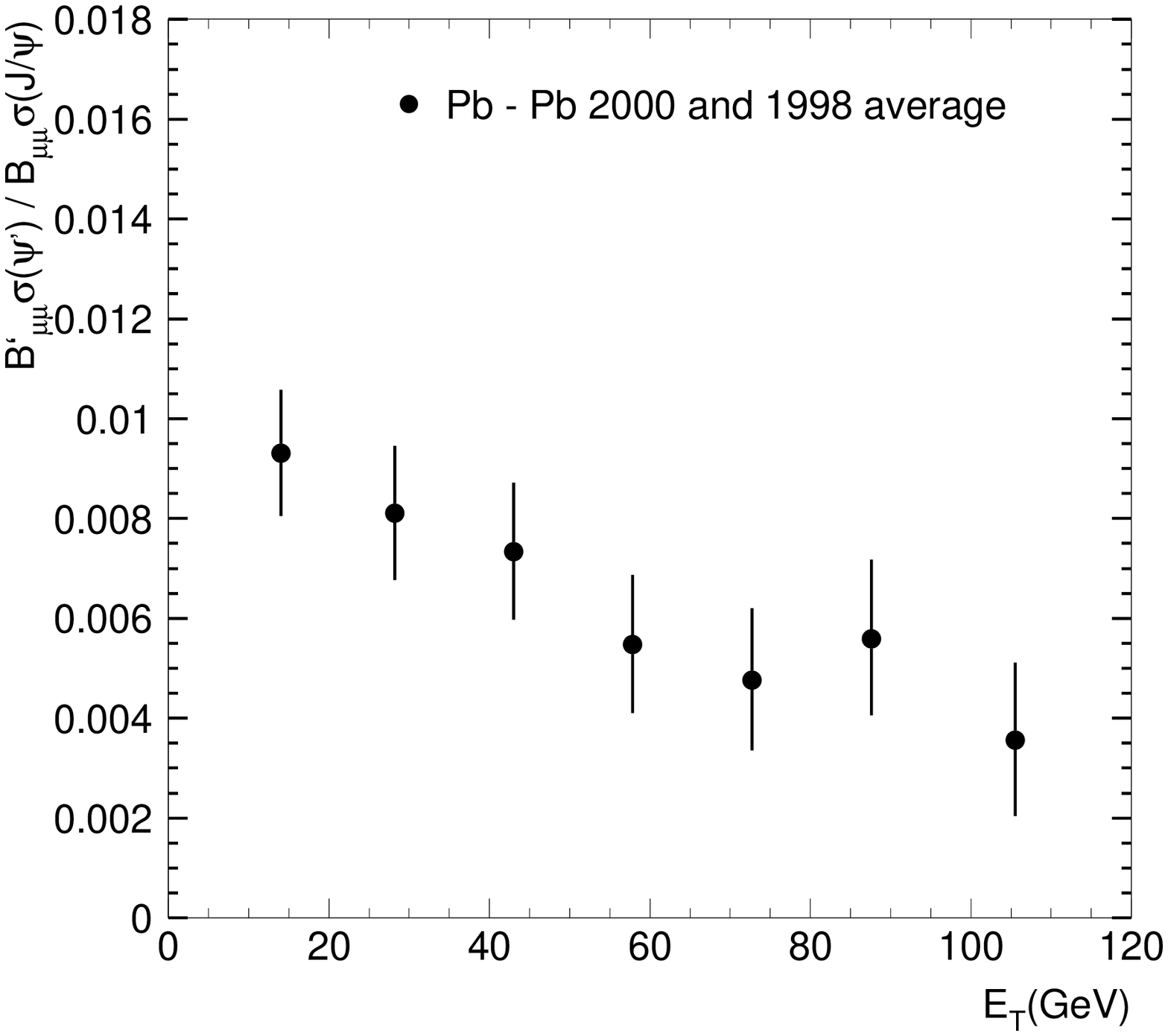}
\end{minipage}
\end{center}
  \vspace*{-15mm}
\hspace*{-15mm}
 \caption{\small\label{psippsiet}$B'_{\mu\mu}\sigma(\psi')/B_{\mu\mu}\sigma($J$/\psi)$
 as a function of $E_T$ for the Pb-Pb 1998 and 2000 data samples, separately for each year (left) and after combining both data sets (right). Errors are the quadratic sum of statistical and systematic uncertainties.}
\end{figure}

The study of the $\psi'$ suppression as a function of centrality is based on 380 
and 905 $\psi'$ events collected in 1998 and 2000, respectively, and analyzed as a function of their transverse energy used as the centrality estimator. Preliminary results of these studies have already been presented in ref.~\cite{QM2004_psip}.\\
\indent
For each centrality class separately, the analysis is performed as described in section~\ref{Experimental setup, data selection and analysis method}. 
In order to get the production pattern as a function of centrality,
results in different classes have to be normalized with respect to each other. We have chosen to make use of the number of elementary nucleon-nucleon collisions for this normalization purpose and have replaced it by a proportional
experimental direct estimate, namely the Drell-Yan cross-section in the
corresponding centrality bin as determined from the same fit procedure. 
 The method thus increases the robustness of the measured centrality pattern,
 as corrections and efficiencies depending on centrality, if any, cancel
 out in the ratio. \\          
\begin{figure}[ht]
\begin{center}
 \begin{minipage}[h]{.47\linewidth}
 \includegraphics[width=\linewidth]{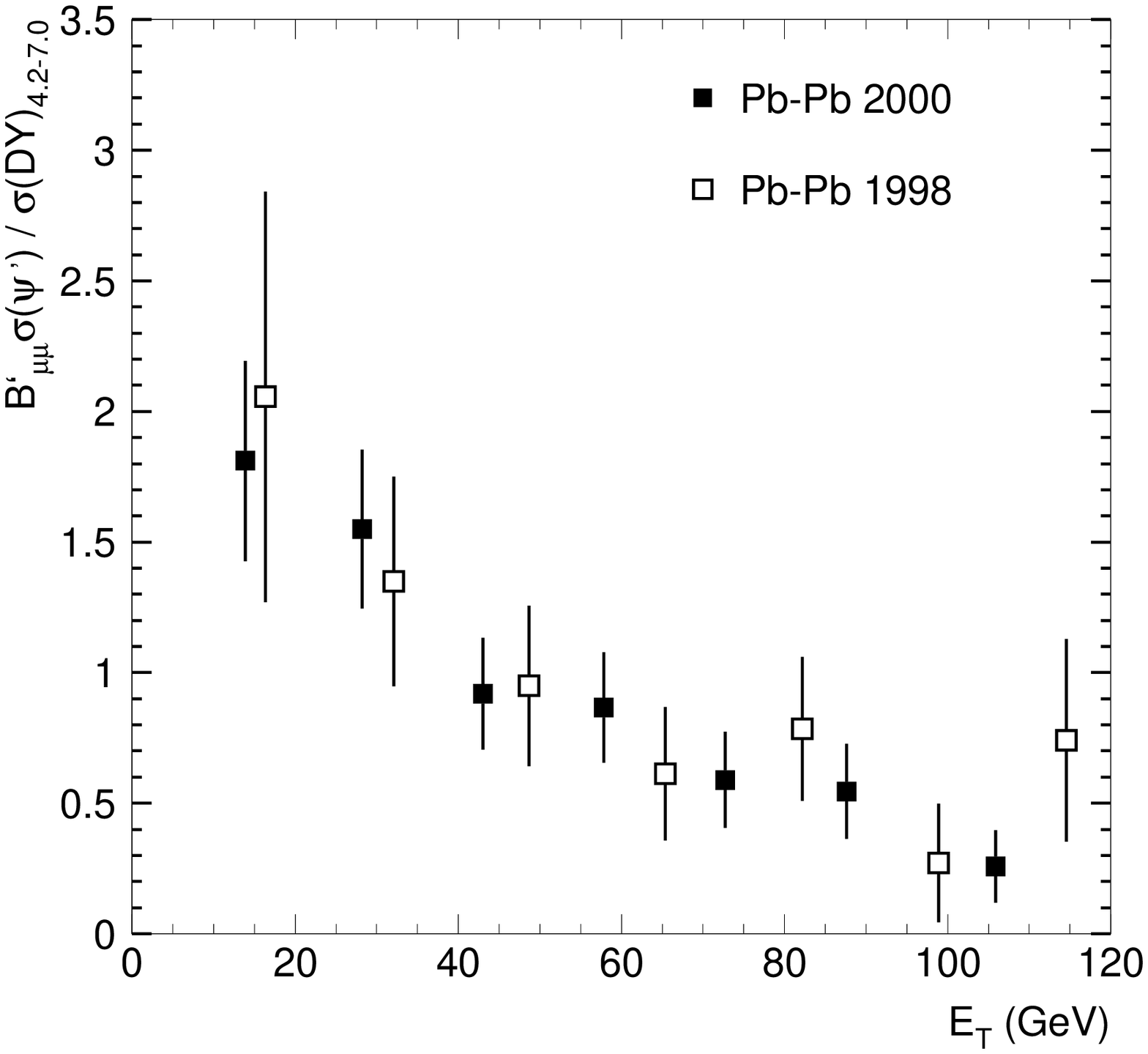}
 \end{minipage} 
\hspace*{6mm}
 \begin{minipage}[h]{.47\linewidth}
  \includegraphics[width=\linewidth]{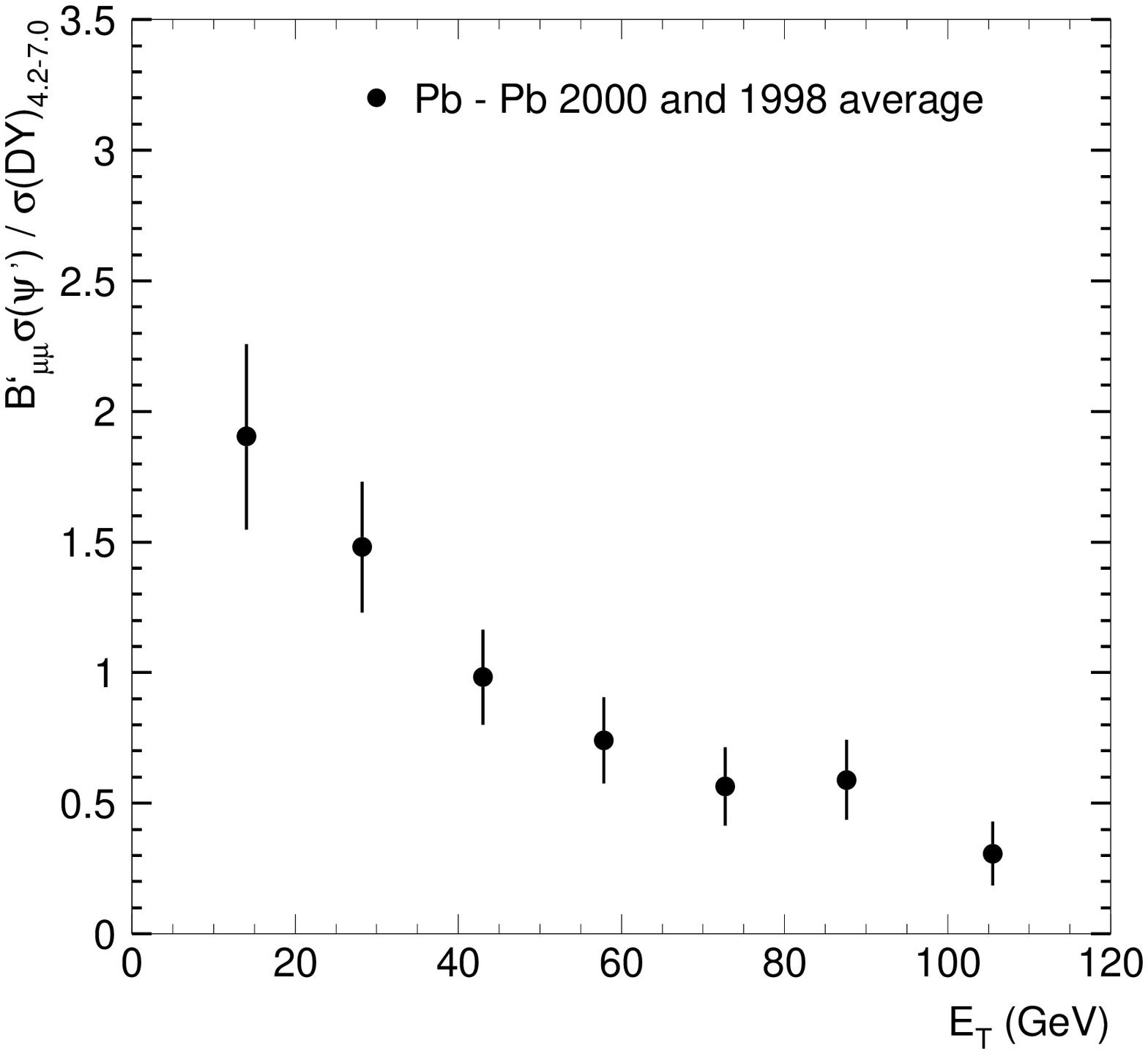}
\end{minipage}
\end{center}
  \vspace*{-15mm}
\hspace*{-15mm}
 \caption{\small\label{psipdyet}$B'_{\mu\mu}\sigma(\psi')/\sigma($DY$)$
 as a function
 of $E_T$  for the Pb-Pb 1998 and 2000 data samples (left). The cross-section
 ratios after combining the results obtained separately (right). Errors are the quadratic sum of statistical and systematic uncertainties.}
\end{figure}
\indent
Figure~\ref{psipdyet} displays the ratio $B'_{\mu\mu}\sigma(\psi')/\sigma($DY$)$
as a function of the transverse energy, for the two data samples.
We should note here that, although the target region was not in vaccum during
the 1998 data taking period, the use of the multiplicity detector to tag in-target interactions has now
allowed the study of peripheral collisions without
ambiguities and, moreover, the results obtained from the year 1998 data sample are 
perfectly consistent with those obtained from the data collected in year
2000, when the target was in vacuum~\cite{hs_thesis}. Thus, the 1998 data have been rebinned to match the same
2000 centrality intervals in order to perform their weighted average. The
combined result shows a suppression of the $\psi'$ production rate which
becomes stronger with increasing centrality, up to a factor of 6 between
peripheral and central collisions. Numerical values are reported in Table~\ref{tab:ratios}. Statistical and systematic errors have been quadratically added. PDF's used to generate Drell-Yan in the Monte Carlo simulations are the main contribution to the systematic errors, as explained in section~\ref{Absolute cross-sections}. They are centrality dependent, but small when compared with the statistical errors. The Drell-Yan mass range chosen to normalize the $\psi'$
 yield is 4.2--7.0 GeV/c$^2$. One can scale up the results to the mass
 range used in the J/$\psi$ suppression studies, 2.9--4.5 GeV/c$^2$~\cite{na50_PbPb2},
 applying the factor 7.96, which is the ratio between the Drell-Yan cross-sections in
 the two mass domains. \\
\indent
The direct comparison between J/$\psi$ and $\psi'$ suppressions as a function of centrality is displayed on Fig.~\ref{psippsiet}, which shows the ratio of cross-sections $B'_{\mu\mu}\sigma(\psi')/B_{\mu\mu}\sigma(J/\psi)$ as a function of transverse energy. The corresponding numerical values are reported in Table~\ref{tab:ratios}. The statistical errors come mainly from the $\psi'$, since the J/$\psi$ statistical uncertainty is less than 1\%. Statistical and systematic errors have been added quadratically. From the steady decrease pattern of this ratio, we conclude that with respect to the J/$\psi$, the $\psi'$  is more and more suppressed with increasing centrality, with a factor 2.5 between peripheral and central reactions.\\
\indent
Table~\ref{tab:events} reports the number of $\psi'$, J$/\psi$ and Drell-Yan fitted events in each transverse energy interval. The numbers refer to both Pb-Pb 1998 and 2000 data taking periods. \\

\begin{table}[*h!]
\begin{center}
\begin{tabular}{||c||c||c||c||}
\hline $E_{T}$ range (GeV)& $\langle E_{T}\rangle$ (GeV)& $B'_{\mu\mu}\sigma(\psi')/\sigma($DY$_{4.2-7.0})$ & $B'_{\mu\mu}\sigma(\psi')/B_{\mu\mu}\sigma($J$/\psi)$ $\times 10^3$ \\
\hline 3--20 & 13.9 & 1.91  $\pm$ 0.35 $^{+0.03}_{-0.06}$ & 9.30 $\pm$ 1.24 $^{+0.28}_{-0.21}$ \\ 
\hline 20--35 & 28.2 & 1.48 $\pm$ 0.25 $^{+0.03}_{-0.04}$ & 8.10 $\pm$ 1.32 $^{+0.30}_{-0.19}$ \\
\hline 35--50 & 43.0 & 0.98 $\pm$ 0.18 $^{+0.02}_{-0.02}$ & 7.33 $\pm$ 1.35 $^{+0.35}_{-0.17}$\\
\hline 50--65 & 57.8 & 0.74 $\pm$ 0.17 $^{+0.02}_{-0.01}$ & 5.48 $\pm$ 1.36 $^{+0.31}_{-0.13}$ \\
\hline 65--80 & 72.7 & 0.56 $\pm$ 0.15 $^{+0.01}_{-0.01}$ & 4.77 $\pm$ 1.40 $^{+0.32}_{-0.11}$ \\
\hline 80--95 & 87.6 & 0.59 $\pm$ 0.15 $^{+0.02}_{-0.01}$ & 5.59 $\pm$ 1.53 $^{+0.43}_{-0.12}$ \\
\hline 95--150 & 105.9 & 0.31 $\pm$ 0.12 $^{+0.01}_{-0.01}$ & 3.56 $\pm$ 1.52 $^{+0.32}_{-0.08}$\\
\hline
\end{tabular}
\end{center}
\caption{\label{tab:ratios}$B'_{\mu\mu}\sigma(\psi')/$DY$_{4.2-7.0}$ and
  $B'_{\mu\mu}\sigma(\psi')/B_{\mu\mu}\sigma($J$/\psi)$ values for each
  centrality bin, for the weighted average between the 2000 and
  1998 results. The first error is statistical, the second systematic.}
\end{table} 

\begin{table}[*h!]
\begin{center}
\begin{tabular}{||c||c||c||c||}
\hline $E_{T}$ range (GeV) & $\psi'$ & J$/\psi$ & $$DY$_{4.2-7.0}$ \\
\hline 3--20 & 186 $\pm$ 25 & 16942  $\pm$ 146 & 112 $\pm$ 8\\ 
\hline 20--35 & 243 $\pm$ 31 & 25229 $\pm$ 181 & 187 $\pm$ 11\\
\hline 35--50 & 227 $\pm$ 35 & 27276 $\pm$ 192 & 264 $\pm$ 12\\
\hline 50--65 & 193 $\pm$ 36 & 27681 $\pm$ 196 & 288 $\pm$ 13\\
\hline 65--80 & 154 $\pm$ 36 & 27315 $\pm$ 200 & 310 $\pm$ 14\\
\hline 80--95 & 159 $\pm$ 37 & 25111 $\pm$ 193 & 311 $\pm$ 14\\
\hline 95--150 & 110 $\pm$ 40 & 28570 $\pm$ 209 & 401 $\pm$ 15\\
\hline
\end{tabular}
\end{center}
\caption{\label{tab:events} The numbers of $\psi'$, J$/\psi$ and DY$_{4.2-7.0}$ fitted events for each centrality range. Errors are only statistical.}
\end{table} 

\indent
\section{Comparison with $\psi'$ production in lighter collision systems}
\label{Comparison with lighter collision systems}
\begin{figure}[!ht]
\begin{center}
 \includegraphics[width=0.6\textwidth]{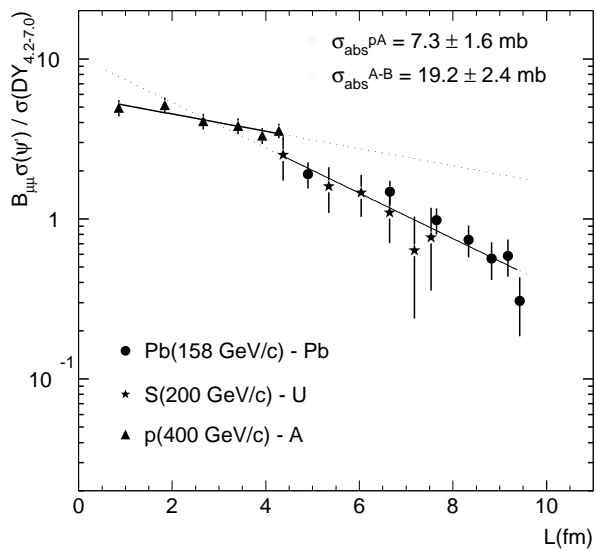}
 \caption{\small\label{psipdyl}$B'_{\mu\mu}\sigma(\psi')/\sigma($DY$)$ as a
 function of $L$. Statistical and systematic uncertainties are added quadratically.}
\end{center} 
\end{figure}

\begin{figure}[!ht]
\begin{center}
 \begin{minipage}[h]{.47\linewidth}
 \includegraphics[width=\linewidth]{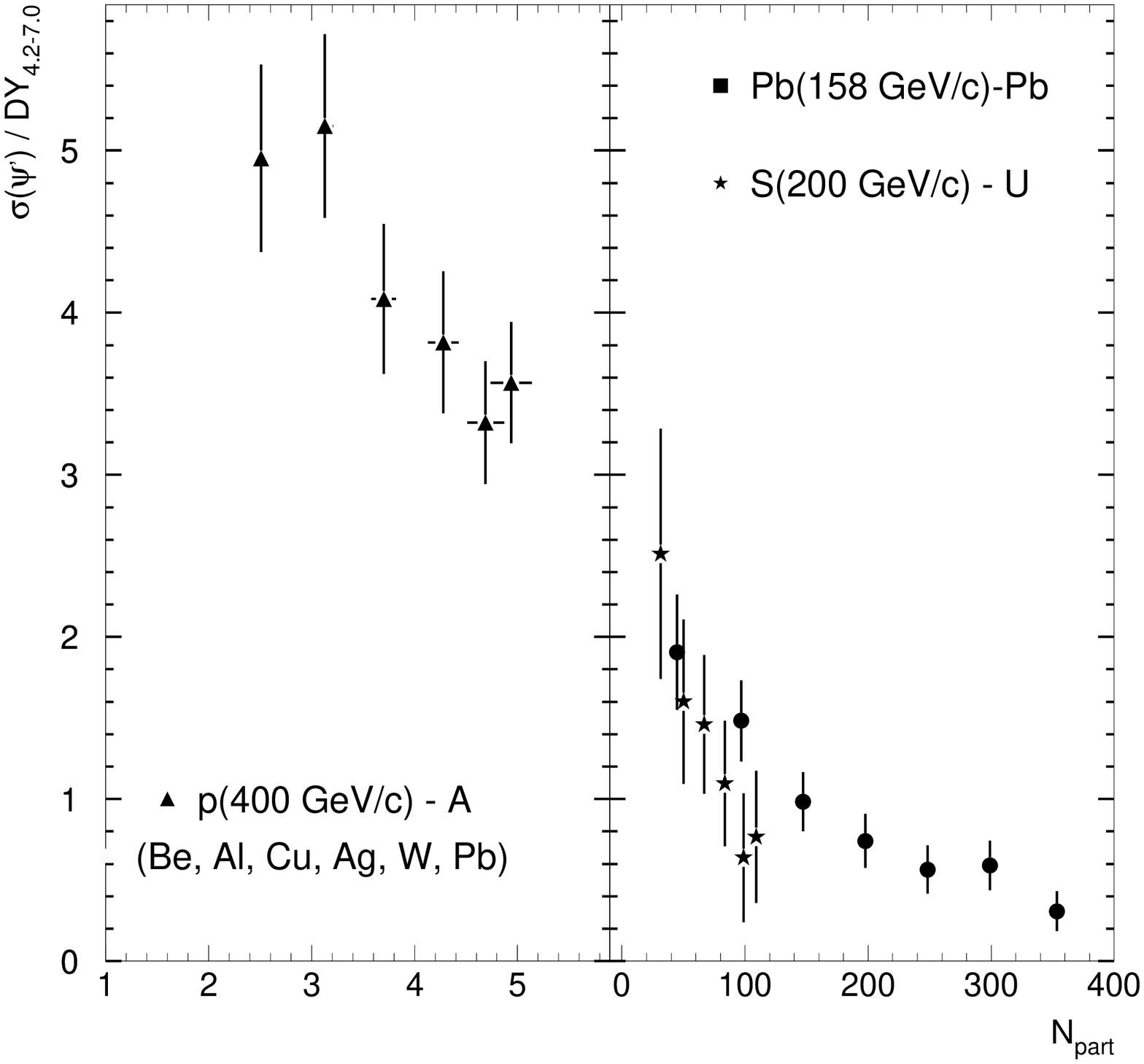}
 \end{minipage} 
\hspace*{6mm}
 \begin{minipage}[h]{.47\linewidth}
  \includegraphics[width=\linewidth]{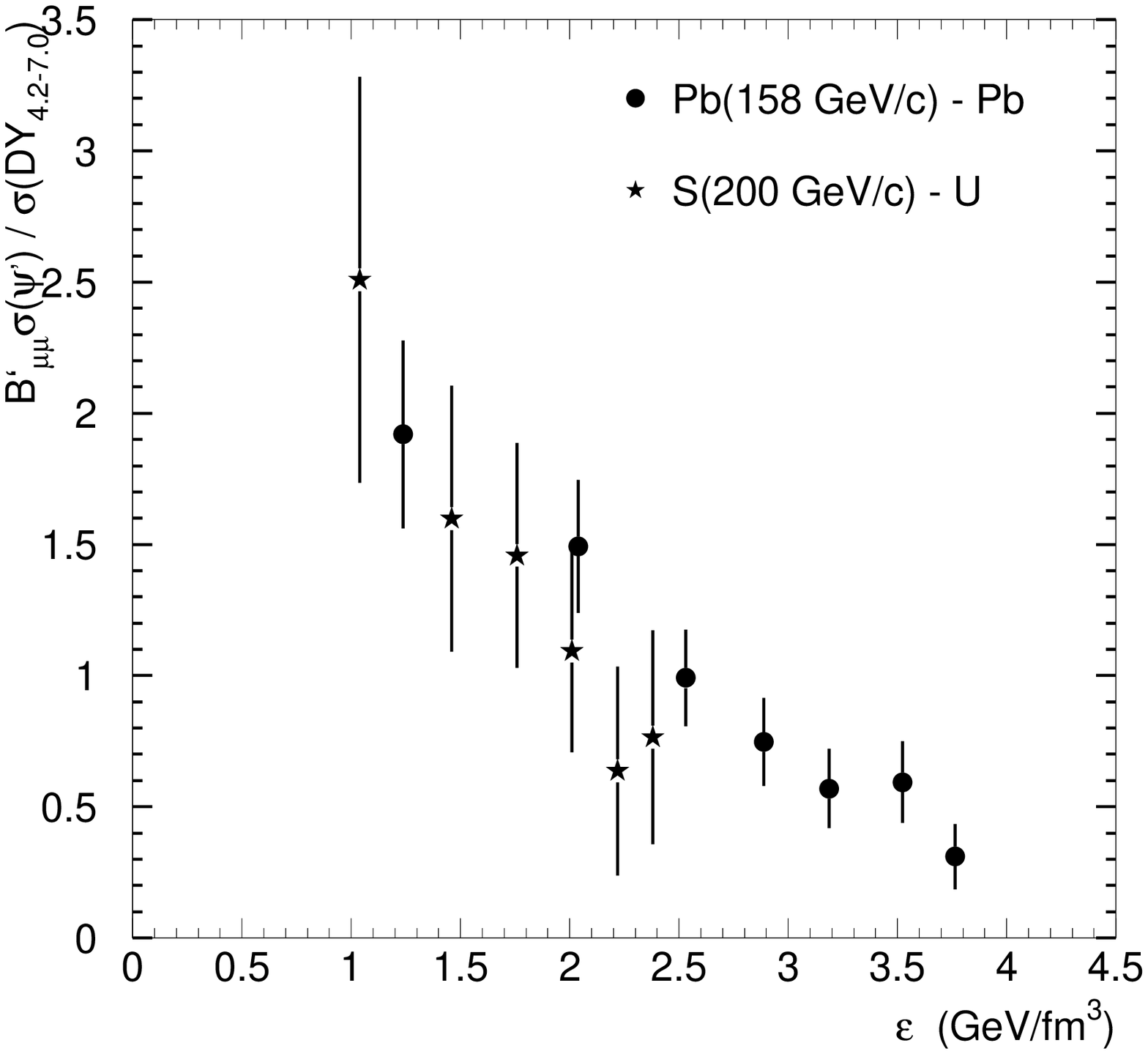}
\end{minipage}
\end{center}
  \vspace*{-15mm}
\hspace*{-15mm}
 \caption{\small\label{psipdynpart}$B'_{\mu\mu}\sigma(\psi')/\sigma($DY$)$ as
 functions of the number of participating nucleons (left) and of the energy
 density (right). Statistical and systematic uncertainties are added quadratically.}
\end{figure}

$\psi'$ suppression in Pb-Pb collisions can be compared with results obtained both from
NA50 proton-induced and NA38 S-U reactions, the latter reanalysed by the NA50 collaboration~\cite{Goncalo_thesis}. 
Data were collected with incident momentum beams of 400 and 450 GeV/c for protons and of 200~AGeV/c for sulphur ions. 
In the framework of NA50, a coherent comparison can be done, since these data were collected 
with the same spectrometer, so that systematic effects related to trigger and
reconstruction efficiencies are similar for all data sets. Besides, the
transverse energy released in S-U collisions has been measured with a very similar
electromagnetic calorimeter. Correction factors have
been applied to scale down the p-A and S-U center of mass energies and rapidity domains to the Pb-Pb 
conditions~\cite{Goncalo_hardprobes}. Taking into account the relative amount of
protons and neutrons in each colliding nucleus, isospin corrections are
needed when the Drell-Yan cross-section is compared among different
systems. For this reason the measured Drell-Yan cross-sections have been
normalized to proton-proton collisions. One should notice that, since
$\psi'$ production in the NA38/NA50 $x_F$ domain is dominated by gluon fusion, rather than by quark-antiquark annihilation, the $\psi'$ isospin corrections are negligible.\\
\indent
In order to compare the $\psi'$ behaviour between different colliding systems we consider hereafter three variables directly related to the impact parameter of the interactions, namely $L$, the average path crossed by
the $c\bar{c}$ pair inside the nucleus, $N_{part}$, the number of
nucleons participating in the interaction, and $\epsilon$, the reached energy
density.  They have been calculated
through the Glauber model formalism~\cite{Glauber} of nucleus-nucleus
collisions. Such a model considers an A-B interaction as a superposition of independent interactions between the beam and target nucleons. The model also assumes that the nucleon-nucleon cross section remains unchanged as the two nuclei cross each other. One should bear in mind that this is an approximation because a nucleon, after a collision, may become excited and subsequently interact with other nucleon with a different cross section. The Pb nuclear density is parameterised by a two-parameters Fermi function~\cite{vries}, also known as the Woods-Saxon distribution, $\rho(r)=\rho_0/(1+\exp((r-r_0)/c))$, where $\rho_0$ is the average nuclear density, taken as 0.17 fm$^{-3}$, $r_0$ is the half-density radius of 6.624 fm and $c$ is a diffuseness parameter of 0.549 fm for protons. Taken into account the ``neutron halo'' effect~\cite{Trzcinska, na50_pA3} a different diffuseness parameter (0.667 fm) is used to describe the neutrons distribution inside the Pb nucleus. The same model has been applied to the U nucleus, with a half-density radius of 6.8054 fm and diffuseness parameters of 0.605 fm for protons and 0.786 fm for neutrons. The actual shape of the U nucleus is not taken into account.
Figure~\ref{psipdyl} shows the ratio $B'_{\mu\mu}\sigma(\psi')/\sigma($DY$)$ 
as a function of $L$.  
The measured suppression patterns suggest the following three features: 
a) a fair agreement with exponential behaviours;
b) two different regimes, one for proton and a different one for ion-induced reactions;
c) a similar centrality dependence for S-U and Pb-Pb interations.
Using an exponential parametrization to describe $\psi'$ absorption
as a function of $L$~\cite{L_ref}, according to $\exp(-\langle\rho
L\rangle\sigma_{\rm abs})$, 
the fit of the data gives an absorption cross-section of 7.3~$\pm$~1.6 mb 
in p-A collisions, while 
a much higher value, 19.2~$\pm$~2.4~mb, is obtained for ion-ion collisions
(S-U and Pb-Pb fitted simultaneously).
The left panel of Fig.~\ref{psipdynpart} shows the same ratio as a function
of the number of participants. By definition, a nucleon is designated as
participant if it has at least one inelastic interaction with one or more
surrounding nucleons. For this work the calculation of $N_{part}$ has
followed eq.~9 of Ref.~\cite{klns}.
 It is worth mentioning that, at SPS energies, several
experiments have observed that the number of participating nucleons in a nucleus-nucleus collision scales with the transverse energy~\cite{wa98_sca, detdeta_sca}. The $\psi'$ suppression as a function of the energy
density is shown in the right panel of Fig.~\ref{psipdynpart}, but only for
A-B collisions. In the framework of the Bjorken model~\cite{Bjorken}, the
energy density reached in a nucleus--nucleus collision is proportional to the
number of participating nucleons per unit transverse area (see
Refs.~\cite{Blaiz_Olli, hs_thesis} for examples of such an evaluation). An
initial formation time, $\tau_0$ = 1 fm/c, has been assumed.
It is important to keep in mind that the three centrality variables,
$L,~N_{part}$ and $~\epsilon$, are strongly correlated, since they are all
calculated by the same Glauber model of nucleus--nucleus collisions. Their average and r.m.s. values for each centrality bin are quoted in Tables~\ref{table:centrality_Pb-Pb} and~\ref{table:centrality_S-U}, for Pb-Pb and S-U
colliding systems, respectively. These values take into account the smearing due to the experimental resolution of the electromagnetic calorimeters used by NA50 and NA38. 
Concerning the energy density values, we are quite confident that
      the comparison between S-U and Pb-Pb is robust, while the
      absolute values could suffer from a 20--25\% systematic uncertainty.\\
\indent
In principle, the use of several centrality estimators in the study of the $\psi'$ suppression pattern in two significantly different colliding systems, such as S-U and Pb-Pb,
 should indicate the centrality variable as a function of
 which the two measured suppression patterns would best overlap.
 In practice, however, the statistical errors of the $\psi'$ suppression patterns presently available do not allow us to make very clear statements in this respect, apart from saying that the use of the $N_{part}$ variable seems to result in a somewhat worse overlap between the two data sets. The Pb-Pb and S-U slopes, from a linear fit to the $\psi'$/DY ratio vs. $N_{part}$  or $\epsilon$ in the overlapping centrality range, differ by 1.6 (resp. 1.0) standard deviations. These observations could also be slightly affected by the approximate nature of our calculations in what concerns the shape of the U nucleus, as mentioned before. 
\begin{table}[h!]
\centering
\begin{tabular}{||c||c||c|c||c|c||c|c||}
\hline $E_{T}$ range & $\langle E_{T}\rangle$ & $\langle L\rangle$ &
r.m.s & $\langle N_{part}\rangle$ & r.m.s. & $\langle\epsilon\rangle$ & r.m.s.\\
(GeV) & (GeV) & (fm) & (fm) & & & (GeV/fm$^3$) & (GeV/fm$^3$)\\
\hline 3--20 & 13.9 & 4.90 & 0.84 & 44.6 & 17.7 & 1.24 & 0.37\\ 
\hline 20--35 & 28.2 &6.65 & 0.44 & 96.7 & 18.0 & 2.04 & 0.20\\
\hline 35--50 & 43.0 &7.65 & 0.31 & 147.1 & 19.3 & 2.53 & 0.14\\
\hline 50--65 & 57.8 &8.34 &0.24 & 197.7 & 20.6 & 2.89 & 0.11 \\
\hline 65--80 & 72.7 &8.83 & 0.18 & 248.6 & 22.0 & 3.19 & 0.09\\
\hline 80--95 & 87.6 &9.17 &0.14 & 299.2 & 23.0 & 3.52 & 0.07\\
\hline 95--150 & 105.9 &9.43 & 0.10 & 353.3 & 22.4 &3.76 & 0.06\\
\hline
\end{tabular}
\caption{\label{table:centrality_Pb-Pb}Mean free path crossed by the
  $c\bar{c}$ pair inside the nucleus, number of parti\-cipating nucleons and energy
  densities for each Pb-Pb $E_{T}$ range.}
\end{table} 

\begin{table}[h!]
\centering
\begin{tabular}{||c||c||c|c||c|c||c|c||}
\hline $E_{T}$ range & $\langle E_{T}\rangle$ & $\langle L\rangle$ &
r.m.s & $\langle N_{part}\rangle$ & r.m.s. & $\langle\epsilon\rangle$ & r.m.s.\\
(GeV) & (GeV) & (fm) & (fm) & & & (GeV/fm$^3$) & (GeV/fm$^3$)\\
\hline 13--28 & 22.1 & 4.37 & 0.54 & 31.3 & 9.1 & 1.04 &        0.22\\ 
\hline 28--40 & 34.3 & 5.35 & 0.43 & 50.0 & 9.7 & 1.46 &        0.18\\
\hline 40--52 & 46.3 & 6.04 & 0.41 & 66.7 & 10.8 & 1.76 & 0.16\\
\hline 52--64 & 58.3 & 6.65 & 0.41 & 83.5 & 11.7 & 2.01 & 0.15\\
\hline 64--76 & 70.2 & 7.17 & 0.38 & 98.8 & 11.0 & 2.22 & 0.13\\
\hline 76--88 & 81.7 & 7.53 & 0.29 & 109.0 & 8.2 & 2.38 & 0.09\\
\hline
\end{tabular}
\caption{\label{table:centrality_S-U} Same as previous table, for
  S-U collisions.}
\end{table} 

\section{Comparison between J/$\psi$ and $\psi'$ suppressions}
\label{Comparison between J/psi and psip suppressions}
\begin{figure}[!ht]
\begin{center}
 \includegraphics[width=0.6\textwidth]{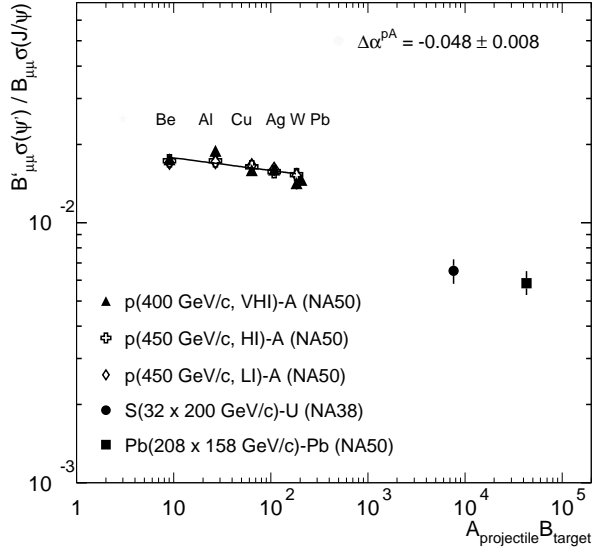}
 \caption{\small\label{psippsiAB}$B'_{\mu\mu}\sigma(\psi')/B_{\mu\mu}\sigma($J$/\psi)$
 as a function of A$_{\rm projectile}\times$B$_{\rm target}$. Statistical and systematic uncertainties are added quadratically.}
\end{center}
\end{figure}

\begin{figure}[!ht]
\begin{center}
\includegraphics[width=0.6\textwidth]{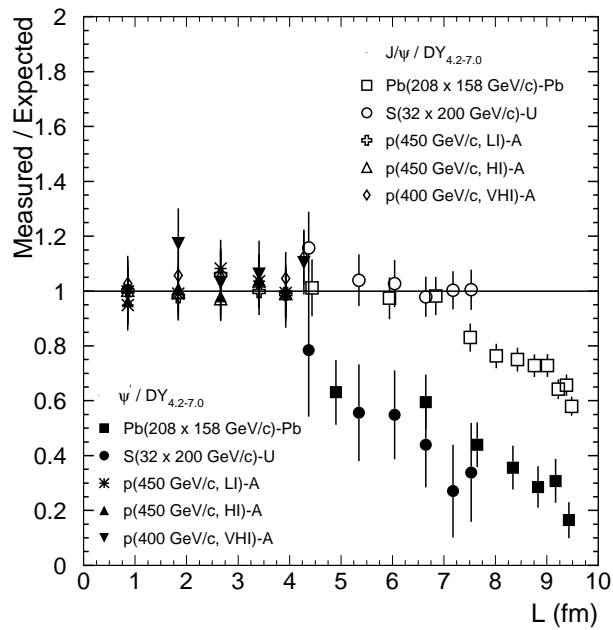}
\caption{\small\label{prop}
The ratio ``measured over expected'' for the relative yields $B'_{\mu\mu}\sigma(\psi')/\sigma($DY$)$ and $B_{\mu\mu}\sigma$(J/$\psi$)$/\sigma($DY$)$, as a function of $L$.}
\end{center}
\end{figure}

\indent
The ratio $B'_{\mu\mu}\sigma(\psi')/B_{\mu\mu}\sigma($J$/\psi)$ as plotted in
Fig.~\ref{psippsiAB} shows the relative suppression of the two
charmonium states as a function of the product of the projectile and
target mass numbers. The ratio is parametrized, in the range of the
p-A data points, by the power law
A$^{\Delta \alpha}$, where $\alpha$ accounts for all nuclear effects
suffered by the resonances. The measured negative value of this
parameter, ${\Delta \alpha}= -0.048~\pm~0.008$, shows
that the $\psi'$ is more absorbed than the J/$\psi$, already in p-A
collisions~\cite{na50_pA2}. Furthermore, a strong suppression of
the $\psi'$\/ with respect to the J/$\psi$ is measured for the ion induced reactions, S-U and Pb-Pb.\\
\indent
Figure~\ref{prop} shows, for various interacting systems, the ratio
between the measured charmonium yields, normalized to Drell-Yan dimuons,
and the corresponding expected ``normal nuclear absorptions''. The latter are 
computed from the p-A results. Using full Glauber calculations, the absorption cross-sections in nuclear matter are 
$\sigma^{\rm Glb}_{\rm abs}($J$/\psi)$ = 4.2~$\pm$~0.5 mb and 
$\sigma^{\rm Glb}_{\rm abs}(\psi')$ = 7.7 $\pm$ 0.9 mb, respectively~\cite{Goncalo_thesis, na50_pA3}. One should notice the good compatibility between this $\sigma^{\rm Glb}_{\rm abs}(\psi')$ value and the one obtained with a simple exponential in $L$, 7.3~$\pm$~1.6 mb, as reported in section~\ref{Comparison with lighter collision systems}. The figure
shows that the departure from ordinary nuclear absorption occurs at a lower
value of $L$ for $\psi'$ than for J/$\psi$. Moreover, while the
$\psi'$/DY behaviour changes between p-A and the whole centrality range of
ion-ion interactions, the J/$\psi$/DY trend in Pb-Pb collisions begins to be compatible with both p-A and S-U results, but deviates more and more as the collision centrality increases. More details on the J/$\psi$ anomalous suppression in Pb-Pb interactions can be found in Ref.~\cite{na50_PbPb2}.\\
\indent
\section*{Conclusions}
\label{Conclusions}
We have measured the $\psi'$ absolute cross-section in Pb-Pb
collisions at $\sqrt{s}$ = 17.3 GeV, in the NA50 kinematic domain, $B'_{\mu\mu}\sigma(\psi')$ =
0.136~$\pm$~0.013(stat)$+0.010-0.006$(syst) $\mu$b. The study of the $\psi'$ production as a function of centrality shows a steady decrease from peripheral to central collisions. It leads to a $\psi'$ suppression factor of 6 relative to Drell-Yan production and of 2.5 relative to J/$\psi$. 
The comparison with lighter systems shows that the $\psi'$ is much more suppressed in nucleus-nucleus than in proton-nucleus reactions and that the suppression pattern in S-U and Pb-Pb approximately overlap when studied as a function of L or energy density, while the overlap as a function of the number of participating nucleons seems to be somewhat worse. The comparison between the J/$\psi$ and $\psi'$ suppression patterns shows that the $\psi'$ anomalous suppression sets in for lower values of $L$.\\

\indent
\section*{Acknowledgments}
This work was partially supported by Funda\c{c}\~ao para a Ci\^encia e a Tecnologia -- Portugal.

\end{document}